\documentclass[%
aip,
pof,
amsmath,amssymb,
floatfix,
reprint,%
]{revtex4-1}

\usepackage{graphicx}
\usepackage{dcolumn}
\usepackage{bm}
\usepackage{gensymb}
\usepackage{mathrsfs}
\usepackage{xcolor}

\newcommand{\subrm}[1]{_{\mathrm{#1}}}

\newcommand{\rmdee}{\mathrm{d}}

\newcommand{\calA}{\mathcal{A}}
\newcommand{\calB}{\mathcal{B}}
\newcommand{\calC}{\mathcal{C}}

\newcommand{\calS}{\mathcal{S}}

\newcommand{\scrD}{\mathscr{D}}

\newcommand{\scrS}{\mathscr{S}}

\newcommand{\rhof}{\rho\subrm{f}}
\newcommand{\muf}{\mu\subrm{f}}

\newcommand{\deep}{d\subrm{p}}
\newcommand{\emp}{m\subrm{p}}

\newcommand{\Ftotr}{\less{F}_{\mathrm{total},r}}
\newcommand{\Ftottheta}{\less{F}_{\mathrm{total},\theta}}

\newcommand{\Tv}{T\subrm{v}}

\newcommand{\thetac}{\theta\subrm{c}}

\newcommand{\bUf}{\boldsymbol{U}\subrm{f}}

\newcommand{\bl}{\boldsymbol{n}}

\newcommand{\er}{\boldsymbol{e}_{r}}
\newcommand{\etheta}{\boldsymbol{e}_{\theta}}

\newcommand{\less}[1]{\bar{#1}}

\newcommand{\dccless}[1]{\less{d}\subrm{cc}}

\newcommand{\wLess}{\less{w}}

\newcommand{\rLess}{\less{r}}
\newcommand{\xLess}{\less{x}}
\newcommand{\yLess}{\less{y}}
\newcommand{\rhoLess}{\less{\rho}}
\newcommand{\zetaLess}{\less{\zeta}}

\newcommand{\psiLess}{\less{\psi}}

\newcommand{\tLess}{\less{t}}

\newcommand{\bFLess}{\less{\boldsymbol{F}}}
\newcommand{\bFtotLess}{\bFLess\subrm{total}}

\newcommand{\Ndot}{\dot{N}}

\newcommand{\etaMean}{\langle \eta \rangle}
\newcommand{\etaAmp}{\eta\subrm{a}}

\newcommand{\partialDeriv}[2]{\frac{\partial #1}{\partial #2}}

\usepackage[utf8]{inputenc}
\usepackage[T1]{fontenc}
\usepackage{mathptmx}

\begin{document}

\title[Direct interception or inertial impaction?]{Direct interception or inertial impaction? A theoretical derivation of the efficiency power law for a simple and practical definition of capture modes}

\author{Mouad Boudina}
 \email{mouad.boudina@polymtl.ca.}
\affiliation{%
Mechanical Engineering Department, Polytechnique Montréal, Montréal, QC H3T 1J4, Canada
}%
\affiliation{%
Laboratory for Multiscale Mechanics (LM2), Polytechnique Montréal, Montréal, QC H3T 1J4, Canada
}%

\author{Frédérick P. Gosselin}
\affiliation{%
Mechanical Engineering Department, Polytechnique Montréal, Montréal, QC H3T 1J4, Canada
}%
\affiliation{%
Laboratory for Multiscale Mechanics (LM2), Polytechnique Montréal, Montréal, QC H3T 1J4, Canada
}%

\author{Stéphane Étienne}%
\affiliation{%
Mechanical Engineering Department, Polytechnique Montréal, Montréal, QC H3T 1J4, Canada
}%


\begin{abstract}
We study the capture of particles advected by flows around a fixed cylinder.
We derive theoretically the power law of the capture efficiency, usually obtained from data fitting only.
Simulations of particle trajectories reveal that captured particles following the power law are smaller than the boundary layer of the cylinder and experience direct interception, whereas the ones diverging from it are larger and observe inertial impaction.
We show that a simple comparison between the particle size and boundary layer thickness splits accurately numerical results into their dominant capture mode.
This criterion is practical in experiments and simulations, and would lift the controversy on the scaling of the capture efficiency.
\end{abstract}

\maketitle
             
\section{Introduction}

Capturing particles is ubiquitous in the living world and a vital process by which many species survive. Plants capture pollen grains to reproduce,\cite{vogel_life_1996} corals filter the sea water from plankton to feed,\cite{hopley_encyclopedia_2010} and larvae catch debris with their fans.\cite{chance_hydrodynamics_1986, widahl_flow_1992}
For industrials too, particle capture is crucial to predict ash deposition on coal boilers and estimate fuel prices,\cite{huang_prediction_1996} or design efficient air filtering systems to ensure hygienic and safe conditions for workers, as well as for products, especially in sensitive sites such as food and pharmaceutical firms.\cite{who_hazard, who_guidelines}
The importance of particle capture and the prevalence of its applications \cite{ding_selective_2015, mollicone_particles_2019, dbouk_respiratory_2020} led several researchers to investigate this process and study how it depends on flow and particle parameters.

The problem we study consists of particles advected by a certain flow around a collector. Traditionally, the collector is a circular cylinder and particles are spherical. This choice is convenient since it brings the study back into a familiar fluid dynamics problem and constitutes a basic benchmark for comparison. In addition, the collector is usually considered as a perfect sink, meaning that particles are counted as captured once they touch the collector edge. Some works extend this model and take into account the possibility of particles to detach after interception,\cite{ginn_effects_1992, moran_particle_1993} or even to accumulate around the collector.\cite{adamczyk_kinetics_1984, hewett_transient_2015} 

The quantity evaluating the capture process is the capture efficiency, denoted usually as $\eta$. The definition of $\eta$ is not unique in the literature, though represents globally the fraction of captured particles from those initially launched.\cite{haugen_particle_2010} Instead of calculating the number, an equivalent definition considers the rate of captured particles $\Ndot$ normalised by the rate of initially released ones $\Ndot\subrm{init}$ \cite{weber_interceptional_1983, palmer_observations_2004}
\begin{equation}
\eta = \frac{\Ndot}{\Ndot\subrm{init}}.
\label{eq:original_def_eta}
\end{equation}
The first parameter affecting capture efficiency is the ratio of the particle diameter $\deep$ to the cylinder diameter $D$
\begin{equation}
R = \frac{\deep}{D}.
\end{equation}
Theoretical,\cite{weber_interceptional_1983} experimental,\cite{palmer_observations_2004} and numerical \cite{haugen_particle_2010, espinosa-gayosso_particle_2012, espinosa-gayosso_particle_2013} studies agree that the capture efficiency increases quadratically with the diameter ratio $\eta \sim R^{2}$. The second parameter involved in the capture efficiency is the collector-based Reynolds number
\begin{equation}
Re = \frac{\rhof U_{0} D}{\muf},
\end{equation}
where $U_{0}$, $\rhof$, and $\muf$ are respectively the upstream flow speed, the fluid density, and its dynamic viscosity. Unlike the diameter ratio, there is a large discrepancy regarding the exponent~$n$ in the scaling $\eta \sim Re^{n}$. Whereas water flume experiments of Palmer \textit{et al.} \cite{palmer_observations_2004} and direct numerical simulations (DNS) of Haugen and Kragset \cite{haugen_particle_2010} showed that $\eta \sim Re^{0.7}$, the semi-analytical derivation of the capture efficiency done by Weber and Paddock \cite{weber_interceptional_1983} revealed a smaller exponent $\eta \sim Re^{0.5}$. On the other hand, using two-dimensional and three-dimensional DNS, Espinosa-Gayosso \textit{et al.} \cite{espinosa-gayosso_particle_2013} found that the exponent $n$ is not constant, and rather decreases from 0.7 down to 0.5 depending on the state of the flow. When the flow is two-dimensional, it takes values from $n = 0.7$ for $Re \approx 50$, then decreases for $50 < Re \le 180$ until reaching $n = 0.5$ beyond $Re > 200$, as the flow exhibits three-dimensional features.

One important missing part in the literature is a theoretical derivation of the scaling of the capture efficiency in terms of the Reynolds number $\eta \sim Re^{n}$. A theoretical derivation of a scaling, found either numerically or experimentally, is essential in any research topic because it gives insight on the basic physics of the process and unveils its underlying mechanisms. Haugen and Kragset \cite{haugen_particle_2010} particularly attempted an analytical derivation of the efficiency using boundary layer considerations, but they arrived at the scaling $\eta \sim Re^{0.5}$ instead of $\eta \sim Re^{0.7}$, contradicting their own numerical results.

Perhaps the variety in the scaling $\eta \sim Re^{n}$ from one author to the other is due to the difference in the nature of particle trajectories in each study, as Haugen and Kragset \cite{haugen_particle_2010} noticed from the behaviour of $\eta$ when varying the Stokes number. Each type of particle trajectory brings about a specific mode of particle capture. If a particle follows perfectly the flow streamlines and intercepts the collector because of its non-zero size, then the capture is called a `direct interception', whereas if it deviates from streamlines and hits the collector with important inertia, then the capture is an `inertial impaction'.\cite{rubenstein_mechanisms_1977} Typically, the first case is observed for small or neutrally buoyant particles, and the second case corresponds to large or heavy particles.
A possible step to lift the discrepancy in the exponent $n$ is to mention, in each experiment or simulation, the dominant capture mode associated with a particular variation of the efficiency.
However, this task is unfeasible because the definition of these modes, which is based on particle trajectories and streamlines, is vague and qualitative. Even if we concede that a researcher succeeds in tracking particles and comparing their trajectories with streamlines, there is no quantitative threshold that determines to which extent they should diverge or superimpose to consider capture as inertial impaction or direct interception. Marking limits between each mode will make the classification of particle capture results in the literature easier and clearer. Haugen and Kragset \cite{haugen_particle_2010} proposed for each mode a particular range of Stokes numbers, but these limits depend on the simulation case considered, therefore are hard to generalise.

In this paper we present, for the first time, a theoretical method to demonstrate both the square variation in the diameter ratio and the square root variation in the Reynolds number $\eta \sim R^{2}Re^{1/2}$. While supporting this finding with numerical simulations of particle trajectories, we found that this scaling fits only for particles that are both smaller than the boundary layer and following streamlines. This scaling underestimates the capture efficiency for the particles larger than the boundary layer and carrying more inertia. From this respect, we will show that a simple comparison of the ratio of the particle diameter to the boundary layer thickness with a certain threshold suffices to determine the dominant mode involved in the capture process.

\section{Theoretical derivation}
\label{sec:theory}

\begin{figure*}
\centering
\includegraphics{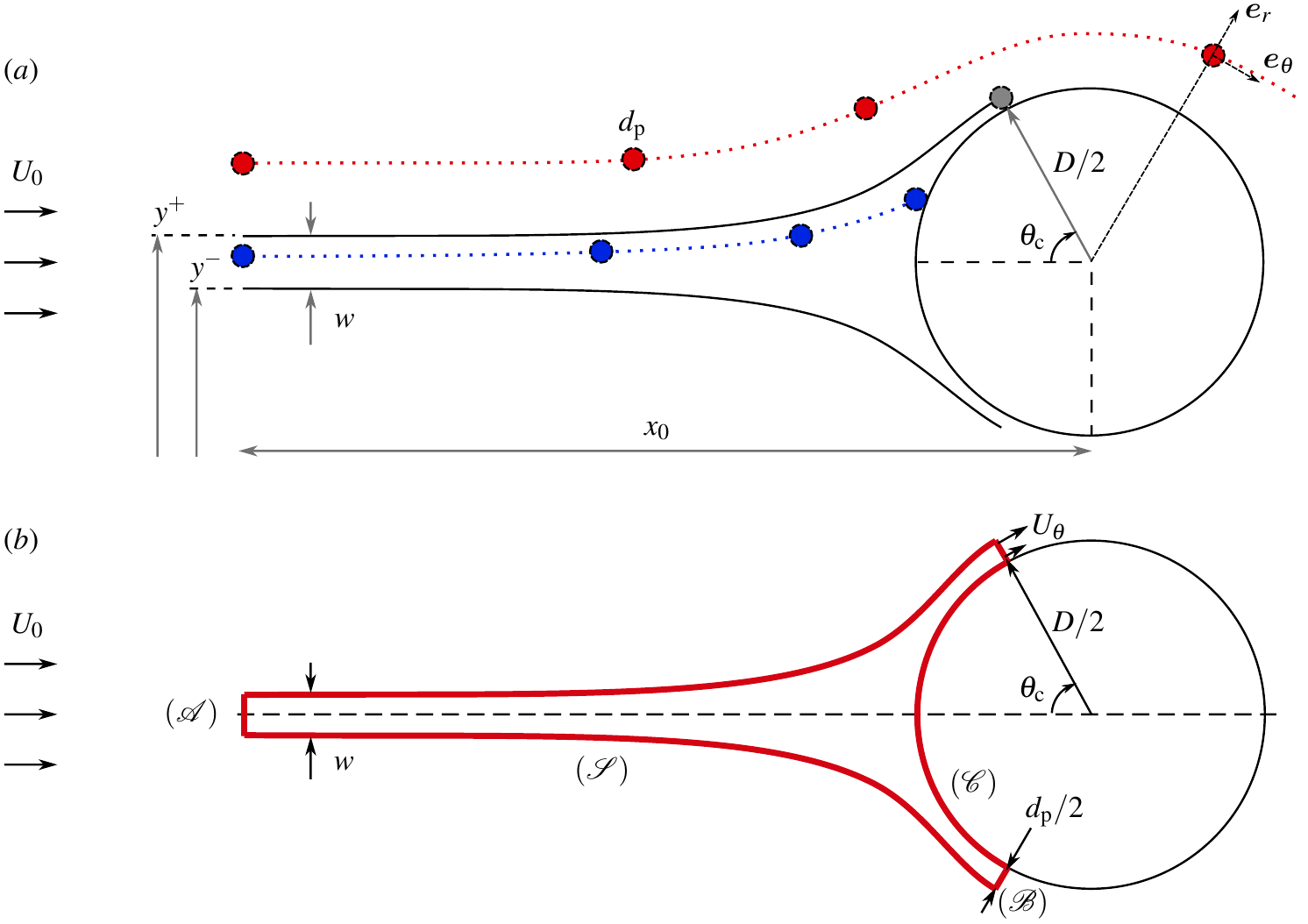}
\caption{($a$) Schematics of particle interception by a fixed cylinder. Particles are launched from a distance $x_{0}$ upstream from the cylinder. Because it starts near the symmetry line, the blue particle is captured, whereas the red particle starts from a large $y$-position and succeeds in crossing the cylinder and escaping capture. The gray particle is the farthest particle that the cylinder intercepts. Its trajectory starts from the ordinate $y^{+}$, and defines the upper border of the capture domain. The size of the capture window is equal to $w = y^{+} - y^{-}$, and $\thetac$ represents the maximum angle of capture.
($b$)~Control volume considered in Eq.~\eqref{eq:balance}.
}
\label{fig:domain_and_balance}
\end{figure*}

We study the interception of particles of diameter $\deep$, advected by a flow of upstream velocity $U_{0}$ around a cylinder of diameter $D$, and positioned at the origin of a polar coordinate system $(r,\theta)$ as illustrated in Fig.~\ref{fig:domain_and_balance}($a$). By definition, the capture rate $\Ndot$ is the number of particles that the cylinder intercepts per unit time. Spreading particles along a starting line from a distance $x_{0}$ prior to the cylinder, there is necessarily an opening around the symmetry line from    which all the ultimately intercepted particles enter. We call this opening the \textit{capture window}, and we denote its size as $w$ (see Fig.~\ref{fig:domain_and_balance}). As a consequence, the rate $\Ndot$ also equals the flux of particles through this capture window 
\begin{equation}
\Ndot =  C_{0}U_{0}w,
\label{eq:def_Ndot}
\end{equation}
where $C_{0}$ is the particle concentration per unit length (\#/m$^{2}$), which we assume constant and uniform.

Owing to the definition \eqref{eq:def_Ndot}, we start our derivation of $\Ndot$ by finding an expression for the size $w$ of the capture window. This latter intervenes in the following balance equation by virtue of mass conservation inside the control volume shown in Fig.~\ref{fig:domain_and_balance}($b$)
\begin{equation}
\int_{(\calA)} \bUf \mathrm{d}\bl
+ \int_{(\calS)} \bUf \mathrm{d}\bl
+ \int_{(\calC)} \bUf \mathrm{d}\bl
+ \int_{(\calB)} \bUf \mathrm{d}\bl = 0,
\label{eq:balance}
\end{equation}
where $\bUf$ denotes the flow velocity. We dropped the fluid density $\rhof$ from Eq. \eqref{eq:balance} because we assume it is constant. The boundaries of the control volume are: the capture window $(\calA)$, the trajectories $(\calS)$ of the outermost captured particles, the cylinder's arc of circle $(\calC)$ defined between the maximum capture angles $\pm \thetac$, and the two segments $(\calB)$ linking the cylinder edge and particle centre. The vector $\rmdee\bl$ is the integration element pointing outwards. In this problem, the angle $\thetac$ is taken as a constant, independent of any flow parameter.

Because of the no-slip condition at the cylinder edge, the integral over $(\calC)$ is equal to zero. Also, assuming that particles follow exactly the streamlines, the integral over $(\calS)$ vanishes. Therefore, we are left with
\begin{equation}
- wU_{0} + 2 \int_{D/2}^{D/2 + \deep/2} U_{\theta} \rmdee r = 0,
\label{eq:definition_e_Utheta}
\end{equation}
with $U_{\theta}$ the tangential component of the fluid velocity. Next, we introduce the stream function $\psi$ through $U_{\theta} = +\partial\psi/\partial r$ (note the `$+$' sign due to the clockwise orientation of the polar angle $\theta$). Then Eq. \eqref{eq:definition_e_Utheta} becomes
\begin{equation}
wU_{0} = 2 \psi(r,\theta) |_{r=D/2 + \deep/2, \theta=\thetac}.
\label{eq:definition-e-psi}
\end{equation}
Now we switch to dimensionless variables
\begin{gather}
\eta = \frac{\Ndot}{C_{0}U_{0}D},\quad
\wLess = \frac{w}{D},\quad
R = \frac{\deep}{D}, \nonumber\\
\psiLess = \frac{\psi}{U_{0}D},\quad
\rLess = \frac{r}{D}.
\end{gather} 
Note that the dimensionless capture rate is exactly the capture efficiency considered in previous studies \cite{weber_interceptional_1983, palmer_observations_2004} and defined in Eq.~\eqref{eq:original_def_eta}, since $C_{0}U_{0}D = \Ndot\subrm{init}$ is the flux of particles initially launched from an opening of size $D$. Thus, from Eqs.~\eqref{eq:def_Ndot} and \eqref{eq:definition-e-psi} we can write
\begin{equation}
\eta = \wLess = 2 \psiLess(\rLess,\theta)|_{\rLess=1/2 + R/2, \theta=\thetac}.
\label{eq:definition_eta_psiLess}
\end{equation}
For very small particles $R \ll 1$, the capture efficiency in Eq.~\eqref{eq:definition_eta_psiLess} can be expressed as a Taylor series around $\rLess = 1/2$
\begin{align}
\eta \approx &\ 2\psiLess(\rLess,\theta)|_{\rLess=1/2,\theta=\thetac}
+ 2\left(\frac{R}{2}\right) \partialDeriv{\psiLess}{\rLess}\bigg\rvert_{\rLess=1/2, \theta=\thetac} \nonumber\\
&+ \left(\frac{R}{2}\right)^{2} \partialDeriv{^{2}\psiLess}{\rLess^{2}} \bigg\rvert_{\rLess=1/2,\theta=\thetac}.
\end{align}
Again, $\partial\psiLess/\partial \rLess \rvert_{\rLess=1/2} = 0$ due to the no-slip condition, and $\psiLess\rvert_{\rLess=1/2} = 0$ because the cylinder edge is a streamline \textit{per se}. We thus obtain the quadratic variation in the diameter ratio of the dimensionless capture efficiency
\begin{equation}
\eta \approx \frac{1}{4} R^{2} \partialDeriv{^{2}\psiLess}{\rLess^{2}} \bigg\rvert_{\rLess=1/2,\theta=\thetac}.
\label{eq:eta_R2_psi}
\end{equation}
This is the stage where previous works ended their theoretical analysis,\cite{weber_interceptional_1983, haugen_particle_2010, espinosa-gayosso_particle_2013} agreeing on the quadratic variation in the diameter ratio. Here we continue our derivation and propose a method to get into the square root variation in the Reynolds number.

Because we assume the particle very small compared to the cylinder, the cylinder wall appears like a flat plate. In this case we can use the analytical expression of the stream function in the boundary layer \cite{schlichting_deceased_boundary-layer_2017}
\begin{equation}
\psiLess = \sqrt{\frac{2\xLess}{Re}} f \left(\yLess \sqrt{ \frac{Re}{2\xLess} } \right),
\label{eq:psi_flat_plate}
\end{equation}
where $Re$ is the cylinder-based Reynolds number, and $f$ is the solution of the Blasius equation. Our approach is to consider a new coordinate system that transforms a flat plate in the $(x,y)$-plane into a circle in the $(\theta, \zeta)$-plane
\begin{subequations}
\begin{align}
x &= x(\theta, \zeta), \\
y &= y(\theta, \zeta).
\end{align}
\label{eq:general_transform}
\end{subequations}
To find the expression of $\psiLess$ in the new $(\theta,\zeta)$-plane, we cannot insert \eqref{eq:general_transform} straightforwardly into \eqref{eq:psi_flat_plate}, because this latter is an inner expansion of the boundary layer solution.\cite{van_dyke_perturbation_1975} Instead, we turn to the Kaplun's correlation theorem,\cite{kaplun_role_1954} which states that if the plate is defined by $(x, y=0)$, and the new wall geometry is defined by $(\theta,\zeta=0)$, then the stream function in the new coordinate system reads
\begin{equation}
\psiLess = \psiLess \left( \xLess(\theta, 0), \zetaLess \partialDeriv{\yLess}{\zetaLess}(\theta, 0) \right).
\label{eq:psi_Kaplun_theorem}
\end{equation}

Let us consider the Joukowski transformation \cite{milne-thomson_theoretical_1962} $\mathscr{J}$
\begin{equation}
z = \mathscr{J}(Z) = Z  + \frac{(D/2)^{2}}{Z},
\label{eq:Joukowski}
\end{equation} 
Putting $Z = (\zeta + D/2) e^{i\theta}$ and $z = x + iy$, the transformation yields
\begin{subequations}
\begin{align}
&x(\theta, \zeta) = \left( \zeta + D/2 + \frac{(D/2)^{2}}{\zeta + D/2} \right) \cos\theta, \\
&y(\theta, \zeta) = \left( \zeta + D/2 - \frac{(D/2)^{2}}{\zeta + D/2} \right) \sin\theta.
\end{align}
\label{eq:transformation}
\end{subequations}

\begin{figure*}
\centering
\includegraphics{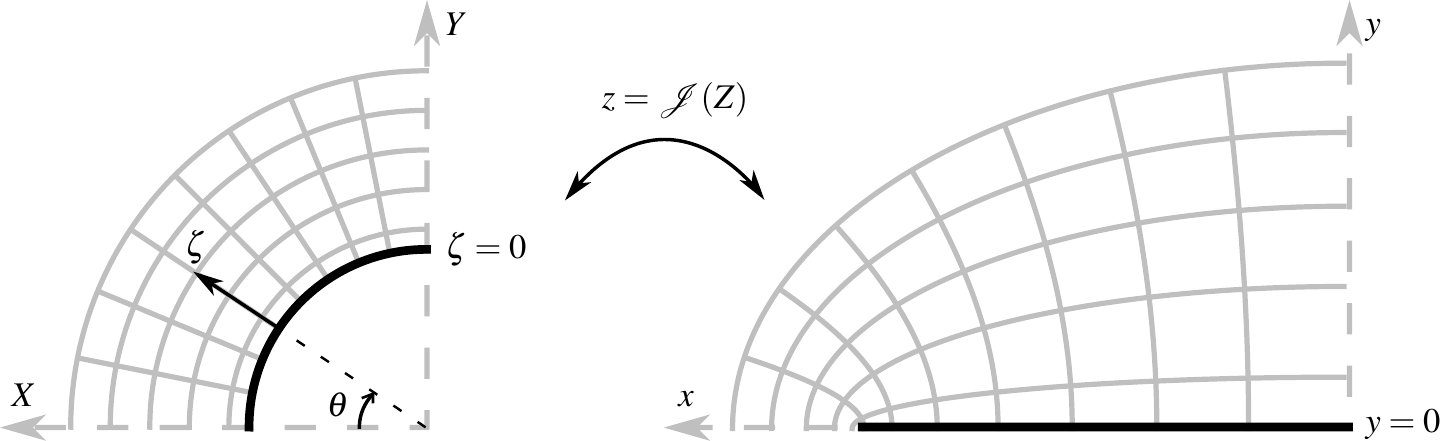}
\caption{Joukowski transformation as defined in Eqs.~\eqref{eq:Joukowski} and \eqref{eq:transformation}. The circle $\zeta = 0$ in the ($X,Y$)-plane corresponds to the flat plate $y = 0$ in the ($x,y$)-plane.}
\label{fig:joukowski}
\end{figure*} 

The Joukowski transformation is shown in Fig.~\ref{fig:joukowski}. This transformation fulfills the condition of Kaplun's theorem since $\zeta = 0$, which represents the circle of radius $D/2$, leads to $y = 0$. From Eq.~\eqref{eq:transformation} we obtain
\begin{subequations}
\begin{align}
x(\theta, \zeta)|_{\zeta=0} &= D \cos\theta, \\
\zeta \partialDeriv{y}{\zeta}(\theta, \zeta)\bigg\rvert_{\zeta=0} &= 2\zeta\sin\theta,
\end{align}
\end{subequations}
which reads in dimensionless variables
\begin{subequations}
\begin{align}
\xLess(\theta, \zetaLess)|_{\zetaLess=0} &= \cos\theta, \\
\zetaLess\partialDeriv{\yLess}{\zetaLess}(\theta, \zetaLess)\bigg\rvert_{\zetaLess=0} &= 2\zetaLess\sin\theta.
\end{align}
\label{eq:expansion_Joukowski}
\end{subequations}
Therefore, replacing Eq.~\eqref{eq:expansion_Joukowski} in \eqref{eq:psi_flat_plate} using \eqref{eq:psi_Kaplun_theorem} we get
\begin{equation}
\psiLess = \sqrt{\frac{2\cos\theta}{Re}} f \left(2\zetaLess\sin\theta \sqrt{ \frac{Re}{2\cos\theta} } \right),
\end{equation}
Finally, coming back to the original variable $\rLess = \zetaLess + 1/2$, and since $\partial^{2}/\partial\rLess^{2} = \partial^{2}/\partial \zetaLess^{2}$, the second derivative of the stream function at $\rLess = 1/2$ (i.e. $\zetaLess = 0$) is
\begin{equation}
\partialDeriv{^{2}\psiLess}{\rLess^{2}}\bigg\rvert_{\rLess=1/2, \theta=\thetac}
= (2\sin\thetac)^{2} \sqrt{ \frac{Re}{2\cos\thetac} } f''(0),
\label{eq:psiLess_squared}
\end{equation}
with $f''(0) = 0.4696$.\cite{schlichting_deceased_boundary-layer_2017, van_dyke_perturbation_1975} Substituting Eq.~\eqref{eq:psiLess_squared} into the capture efficiency in Eq.~\eqref{eq:eta_R2_psi} we have
\begin{equation}
\eta \approx \frac{\sin^{2}\thetac f''(0)}{\sqrt{2\cos\thetac}} R^{2} Re^{1/2}.
\label{eq:power_law}
\end{equation}
We find a capture efficiency which varies with the square of the diameter ratio $R$ and the square root of the Reynolds number $Re$.
Notice that the particle intervenes in the formula of $\eta$ in Eq.~\eqref{eq:power_law} only through its size, whereas its density does not figure in any part of the derivation. This independence is in harmony with the simulations of Espinosa-Gayosso \textit{et al.},\cite{espinosa-gayosso_density-ratio_2015} which yielded capture efficiencies unaffected by the particle density when the Stokes number is very small $Stk \ll 1$.

To validate the scaling in Eq.~\eqref{eq:power_law}, we simulate the advection of particles having diameter ratios $0.008 \le R \le 0.1$ in fluid flows at Reynolds numbers $3 \le Re \le 300$ around a circular collector, following the same methodology described in Boudina \textit{et al.} \cite{boudina_jfm_2020} In brief, we perform two-dimensional DNS with \textsc{Cadyf},\cite{etienne_perspective_2009} an in-house monolithic finite element solver. Then we export the flow solution into \textsc{Paradvect},\cite{mou3adb_paradvect_2020} a Python code that integrates the momentum equation of each particle. This latter is subjected to hydrodynamic drag, pressure load, and added mass force. However, we neglect Brownian motion, and consider non-motile particles, meaning they cannot change actively their trajectories by swimming for instance.
Again, we calculate the capture efficiency via the size of the capture window $\eta = \wLess$, using the method of automated dichotomy.\cite{boudina_jfm_2020} Because of the periodic vortex shedding, the efficiency varies in time. We assume it has the following ansatz
\begin{equation}
\eta(t) = \etaMean
+ \etaAmp \sin \left( \frac{2\pi t}{\Tv} + \varphi \right),
\label{eq:eta_temporal_decomp}
\end{equation}
where $\Tv$ is the vortex shedding period and $\varphi$ the phase.

\begin{figure*}
\centering
\includegraphics{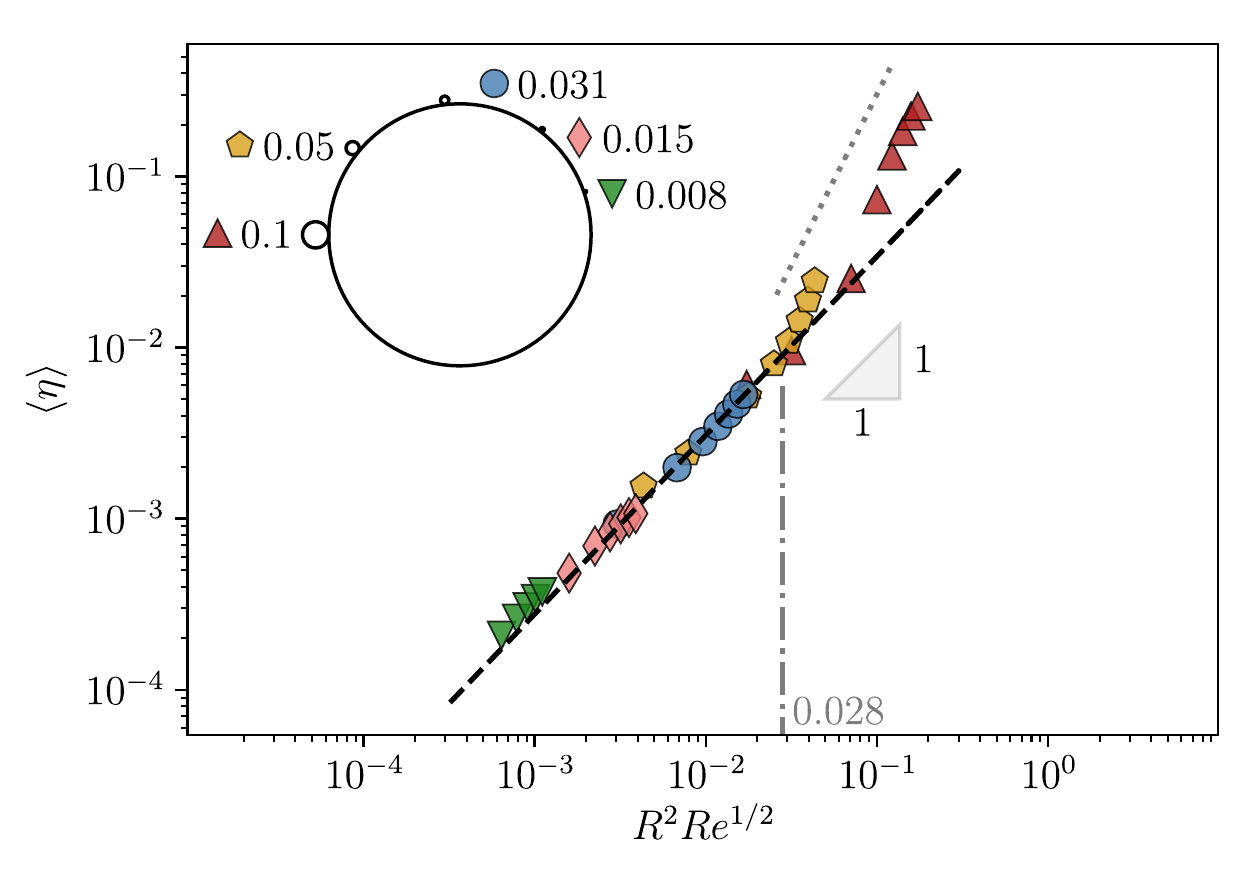}
\caption{The mean capture efficiency for a fixed cylinder versus $R^{2}Re^{1/2}$ for $3 \leq Re \leq 300$ and $0.008 \leq R \leq 0.1$. The black dashed line is the power function fit of data points for $R^{2}Re^{1/2} < 0.028$ ($r^{2} = 0.99$): $\etaMean = c R^{a} Re^{b}$ with $a = 2.09 \pm 0.06$, $b = 0.52 \pm 0.01$ and $c = 0.378 \pm 0.044$. The gray dotted line depicts the variation in $\sim R^{2}Re$.}
\label{fig:eta_vs_R2sqrtRe}
\end{figure*}

We find that $\etaAmp/\etaMean < 0.02$, thus the transient term in \eqref{eq:eta_temporal_decomp} represents a small fluctuation, and we will focus only on the mean term $\etaMean$. Fig.~\ref{fig:eta_vs_R2sqrtRe} shows the variation of $\etaMean$ with the product $R^{2}Re^{1/2}$. We can distinguish two regimes: for $R^{2}Re^{1/2} < 0.028$ the mean capture efficiency conforms to the power law \eqref{eq:power_law}, whereas beyond it diverges towards $\sim R^{2}Re$. In this latter zone, our results agree with the finding of Haugen and Kragset,\cite{haugen_particle_2010} where the efficiency varied linearly with the Stokes number $\eta \sim Stk \sim R^{2}Re$ when $Stk > 1$. 

\section{`Direct interception' versus `Inertial impaction'}
\label{sec:modes}

Because of the quadratic variation in the diameter ratio, the product $R^{2}Re^{1/2}$ stays lower than the threshold 0.028 for small particles $R \le 0.031$ advected by flows of Reynolds number $3 \le Re \le 300$, in which case the mean capture rate verifies $\etaMean \sim R^{2}Re^{1/2}$. This result is consistent with the analytical derivation in section \ref{sec:theory}, recalling that we assumed particles extremely small ($R \ll 1$) and following streamlines.
So what do particles in each regime have in special?

\begin{figure*}
\centering
\includegraphics{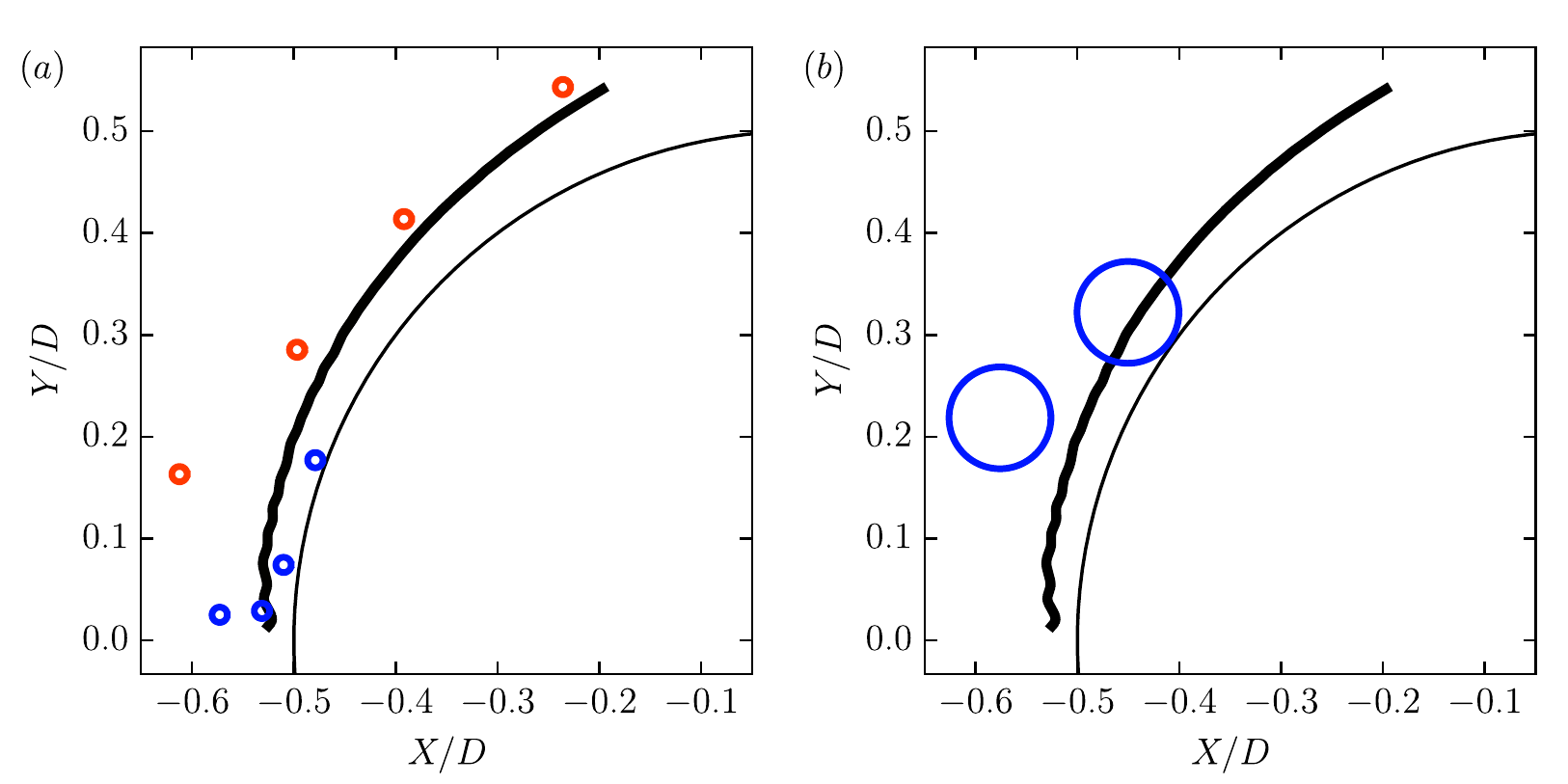}
\caption{Comparison of the numerical boundary layer thickness (bold black line) at $Re = 100$ with particles. For each angle we calculated the distance from the cylinder wall where the angular flow velocity $U_{\theta}$ is equal to 99\% of its value at a faraway position $U_{\theta}|_{r \rightarrow \infty}$. The obtained points are then connected with a cubic spline.
($a$) The blue particle of $R = 0.015$ is small enough to enter inside the boundary layer and remain within until capture. The red particle has the same size but launched far from the separation line. The boundary layer deviates it and prevents the cylinder from interception. ($b$) The particle of $R = 0.1$ is larger than the boundary layer, hence impacts the cylinder unaffected.}
\label{fig:boundary_layer}
\end{figure*}

Fig.~\ref{fig:boundary_layer} contrasts the sizes of the boundary layer thickness of a flow at $Re = 100$ with the particles $R = 0.015$ and 0.1. These cases have a product of $R^{2}Re^{1/2} = 0.002$ and 0.1 respectively. The particle $R = 0.015$ is so small that the boundary layer can entirely englobe it. To hit the cylinder, this particle has no choice but to penetrate the boundary layer. As a result, the boundary layer is an inevitable element in the capture process. On the other hand, the particle $R = 0.1$ is twice larger than the boundary layer, and is able to hit the cylinder from outside. The boundary layer in this case plays no role in the capture process.

\begin{figure*}
\centering
\includegraphics{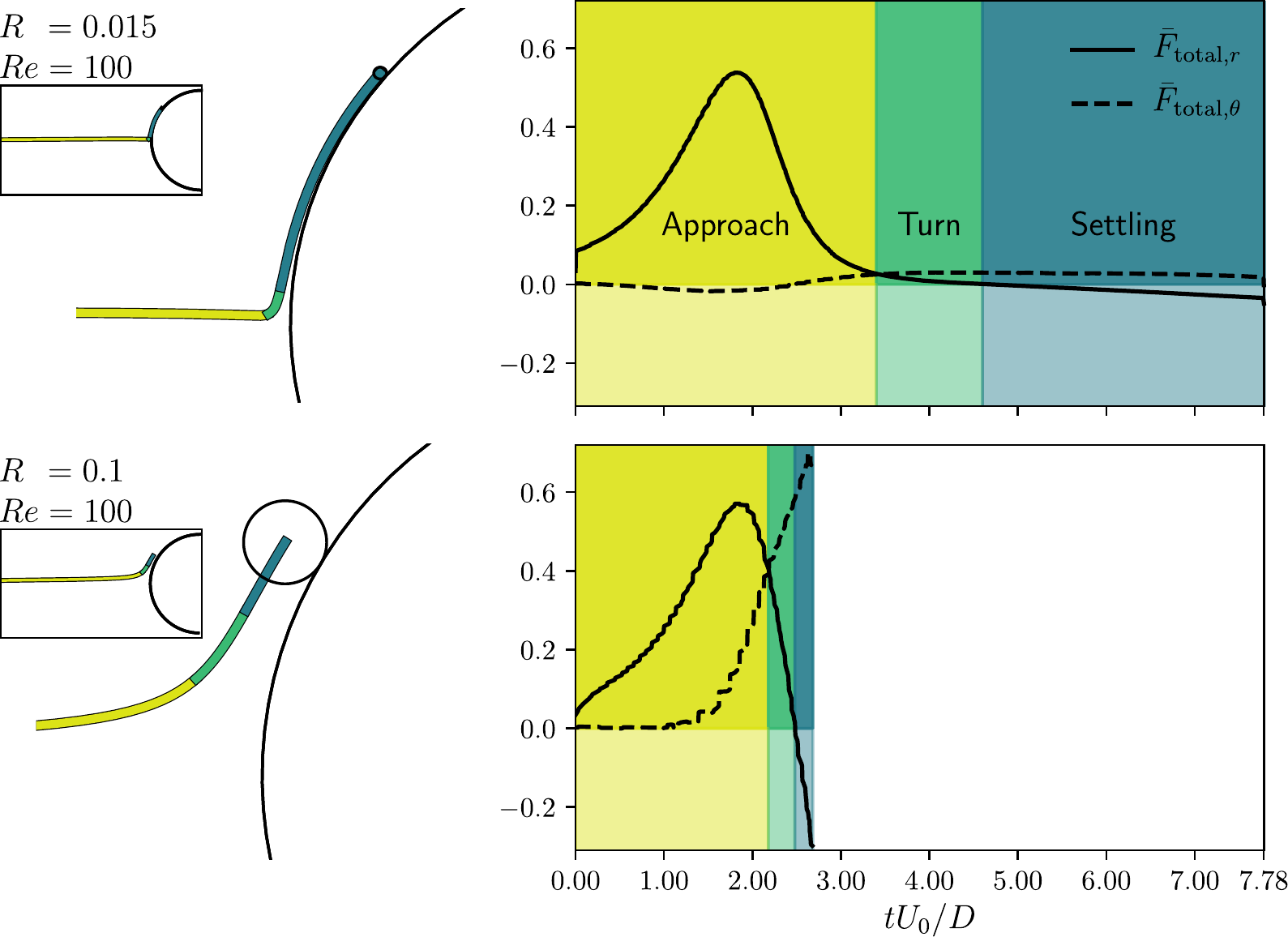}
\caption{Time evolution of the dimensionless radial $\Ftotr$ and angular $\Ftottheta$ components of the total hydrodynamic force applied on the farthest captured particles advected by a flow at $Re = 100$. The radial vector $\er$ points away from the cylinder and the angular vector $\etheta$ is clockwise (see Fig.~\ref{fig:domain_and_balance}($a$) for the polar frame definition). In the case of the particle $R = 0.015$ (top), $\Ftotr$ is positive during the approach phase (in yellow), meaning that it resists the particle from coming to the cylinder. Meanwhile, $\Ftottheta$ is negligible, hence the particle follows a straight path. In the turn phase (in green) we have $\Ftottheta > \Ftotr$, and this latter keeps decreasing until it drops below zero at $\tLess \approx 4.6$. Subsequently, during the settling phase (in blue), the hydrodynamic force brings the particle even closer to the cylinder. However, for $R = 0.1$ (bottom), since $\Ftottheta$ intersects with $\Ftotr$ while this latter is still around its peak, the particle deviates early before getting close to the cylinder. Also, the force components take values of $\Ftotr \approx -0.3$ and $\Ftottheta \approx 0.7$, which is larger than in the previous case $\Ftotr \approx -0.03$ and $\Ftottheta \approx 0.02$.}
\label{fig:trajectory_phases}
\end{figure*}

To identify the behaviour of particle dynamics in this regime, let us examine in Fig.~\ref{fig:trajectory_phases} the time evolution of the total dimensionless hydrodynamic force applied on each of these particles, projected in the radial and angular directions $\Ftotr = \bFtotLess\cdot\er$ and $\Ftottheta = \bFtotLess\cdot\etheta$ (see Fig.~\ref{fig:domain_and_balance}($a$) for the polar frame definition). This force is non-dimensionalised by the quantity $\emp U_{0}^{2}/D$, where $\emp$ is the mass of a single particle. Regarding the small particle $R = 0.015$, we notice that the trajectory comprises three phases. From the starting line, the total force points almost radially away from the cylinder ($\Ftotr > 0$ and $|\Ftottheta|/\Ftotr \ll 1$), meaning that the particle travels almost in a straight line without deviation until it slows down near the cylinder. We term this phase as the \textit{approach}. The next phase, the \textit{turn}, starts when the radial component takes over the angular one, i.e. from the moment when $\Ftottheta \ge \Ftotr$. The fluid then slows down the particle near the stagnation point, then carries it sideways and curves its trajectory. After that, while the particle enters the boundary layer, the radial hydrodynamic force drops below zero and changes its sign $\Ftotr \le 0$. This moment announces the starting of the phase we term as the \textit{settling}. During this phase, the repulsive action of the hydrodynamic load transitions into an attracting action towards the cylinder. While following gently streamlines, the particle drifts and settles down \textit{directly} at the cylinder edge. From this description of the trajectory, we can therefore classify this capture mode as a `direct interception'.

As for the large particle $R = 0.1$, in contrast, we first notice that it moves faster than the smaller one described above, and the durations of the three phases shrink. In the approach phase, $\Ftottheta$ starts climbing up whereas $\Ftotr$ is still important, meaning that the particle deviates early and keeps a considerable radial momentum. We also point out that the trajectory is not sharply curved during the turn, and travels as much distance in it as in the settling phase. Also, at the instant of capture, the magnitude of the radial and angular components of the total force take values $\Ftotr \approx -0.30$ and $\Ftottheta \approx 0.70$, which are 10 and 35-fold greater than in the previous case $\Ftotr \approx -0.03$ and $\Ftottheta \approx 0.02$. Because this particle deviates from streamlines, moves rapidly, and \textit{impacts} the cylinder with high momentum, we deduce that the capture mode in this case is an `inertial impaction'.

Inspired from Fig.~\ref{fig:boundary_layer}, we put forward that the capture mode is a direct interception, hence verifying the scaling $\eta \sim R^{2}Re^{1/2}$, only if the particle diameter $\deep$ is smaller than a fraction $M$ of the boundary layer thickness $\delta$
\begin{equation}
\frac{\deep}{\delta} < M.
\label{eq:condition_DI}
\end{equation}
The constant $M$ is an upper bound that we should determine.

The boundary layer thickness is inversely proportional to the square root of the cylinder-based Reynolds number,\cite{landau_fluid_1966} and we assume that
\begin{equation}
\frac{\delta}{D} \approx \frac{1}{Re^{1/2}}.
\label{eq:def_delta}
\end{equation}
Therefore, we have
\begin{equation}
\frac{\deep}{\delta}
= \frac{\deep}{D} \frac{D}{\delta}
= R Re^{1/2},
\end{equation}
and the direct interception condition becomes equivalent to
\begin{equation}
R Re^{1/2} < M.
\label{eq:condition_DI2}
\end{equation}

\begin{figure*}
\centering
\includegraphics{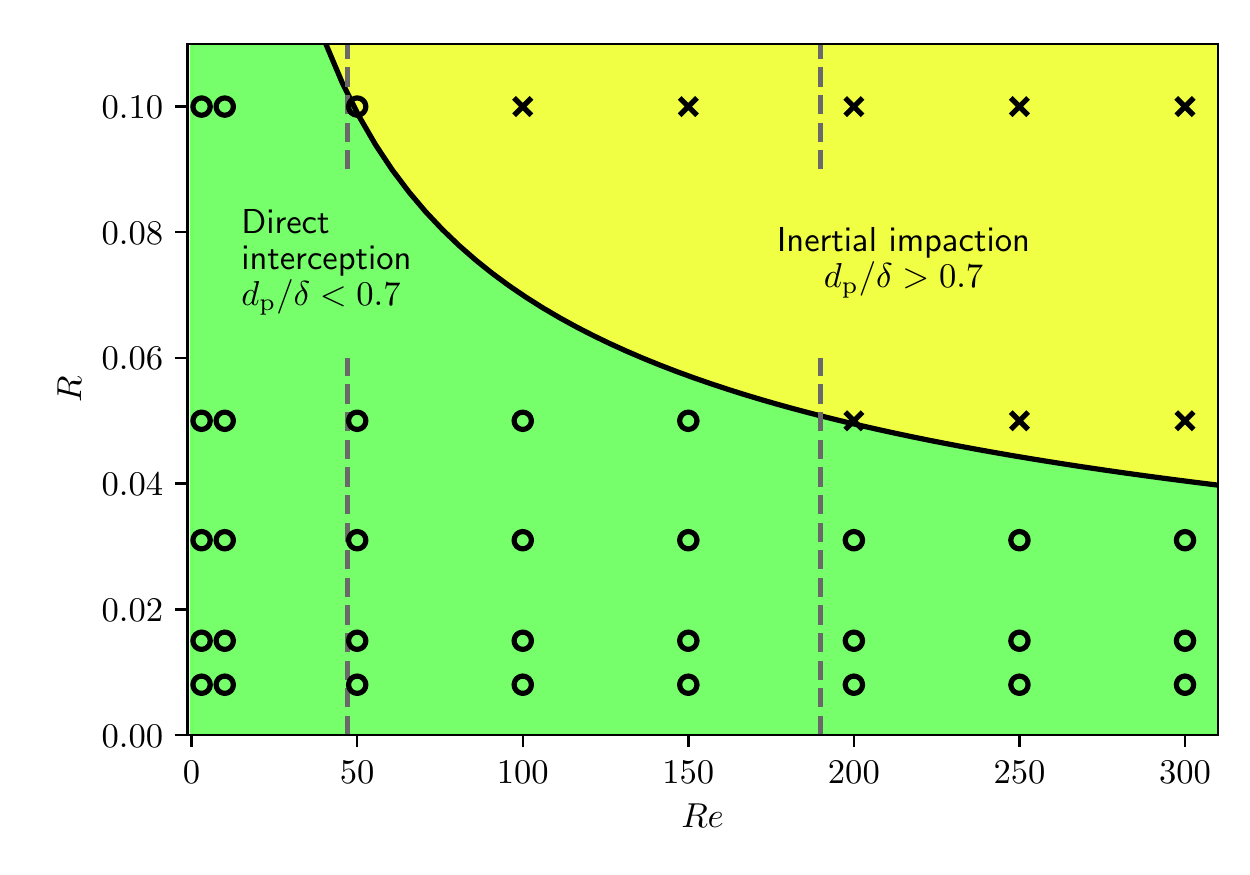}
\caption{Phase diagram of the particle advection simulation cases. The bold black line $R^{*} = 0.7/\sqrt{Re}$ is the separatrix between the direct interception (green) and inertial impaction (yellow) regions. Open circles ($\bigcirc$) indicate simulation cases verifying the scaling $\eta \sim R^{2}Re^{1/2}$, and crosses ($\times$) indicate those that diverge from it. Between the two vertical dashed lines $Re = 47$ and $Re = 180$ is the regime of two-dimensional vortex shedding. Although all simulations performed here are two-dimensional, beyond $Re \ge 180$ the real flow is three-dimensional.}
\label{fig:phase_diagram}
\end{figure*}

We seek the value of $M$ graphically such that the curve $R^{*} = M/Re^{1/2}$ separates numerical data of direct interceptions ($\bigcirc$) and inertial impactions ($\times$) in the phase diagram $(R, Re)$ shown in Fig.~\ref{fig:phase_diagram}. We find that $M \approx 0.7$ is the best value which is consistent with the results of the mean capture rate in Fig.~\ref{fig:eta_vs_R2sqrtRe} and splits well the domain into two distinct regions.
Indeed, we see that small particles $R \le 0.031$ for all Reynolds numbers belong to the region $\deep/\delta < 0.7$ (green), and the larger ones $R = 0.05$ and 0.1 have, respectively, three cases $200 \le Re \le 300$ and five case $100 \le Re \le 300$ belonging to the region $\deep/\delta > 0.7$ (yellow).
More rigorous than a simple graphical argument, we also find the value $M = 0.7$ from a mathematical perspective, which we develop in appendix \ref{app:mathematical}.

Henceforth, in order to score a direct interception, the particle diameter should be smaller than 70\% of the boundary layer thickness
\begin{equation}
\frac{\deep}{\delta} < 0.7,
\end{equation}
which is also equivalent to
\begin{equation}
R Re^{1/2} < 0.7.
\end{equation} 

It is worth mentioning that the separatrix $R^{*} = 0.7/\sqrt{Re}$ being inversely proportional to the square root of the Reynolds number could have been noticed when we derived analytically the expression of $\etaMean$ in section \ref{sec:theory}. Indeed, we dealt with small arguments $\yLess \sim \rhoLess \sqrt{Re}$ of the Blasius solution at a distance from the cylinder wall $\rhoLess \sim R$; in other words $R\sqrt{Re} = \deep/\delta$ was small.

\section{Discussion and conclusion}

For the first time, a complete theoretical derivation of the scaling of the particle capture efficiency of a fixed cylinder $\eta \sim R^{2}Re^{1/2}$ is proposed.
Throughout this derivation, we saw that the exponent 1/2 in the Reynolds number comes from the effect of the boundary layer thickness on the capture process. Indeed, as noted by Haugen and Kragset, \cite{haugen_particle_2010} the boundary layer acts as a `shield' for the cylinder. For low Reynolds numbers, it thickens and protects the cylinder from incident particles, as a result of which the capture is low. For high Reynolds numbers, it becomes thin and weakens in front of the incident particles, whence the capture is high.

We applied the Blasius boundary layer solution to a circular cylinder via the Joukowski conformal transformation. This latter, however, transforms a circle into a finite plate, so one could think of the use of the Blasius solution as debatable since it deals with infinite plates.
In our case, the domain of interest where we calculate the scaling is the frontal region of the cylinder, which corresponds to the tip of the finite plate.
For the small particle, that zone is far from the endpoint of the plate, where we could expect the solution to be substantially different. Therefore we deem the use of the Blasius solution legitimate. Further studies might elaborate on this model by considering other approximations of the laminar boundary layer solution.\cite{howarth_solution_1938}

Additionally, although the flow is in reality periodic due to vortex shedding, we presumed in this derivation that the flow is steady. We think this assumption is reasonable as long as we stay in the frontal region. In fact, streamlines upstream from the cylinder do not deform significantly, so the control volume we considered in Fig.~\ref{fig:domain_and_balance}($b$) remains fixed. Or, from another standpoint, we can view flow variables as time-averaged, as it was the case in the analysis of Espinosa-Gayosso \textit{et al.}\cite{espinosa-gayosso_particle_2013}

Another assumption to revisit concerns the invariance of the maximum angle of capture $\thetac$ with the Reynolds number. The numerical factor we obtained in the power law in Eq. \eqref{eq:power_law} is a function of $\thetac$, which we considered independent of flow variables. Weber and Paddock \cite{weber_interceptional_1983} compiled previous numerical studies and found that the angle $\thetac$ varied between 70$\degree$ at $Re=1$ to 47$\degree$ at $Re=200$. This variation of $\thetac$ with $Re$ could lead to a theoretical exponent for the capture rate in $Re$ different from 1/2. This remains to be explored.

After examining the trajectory behaviour of particles prior to and during capture, we found that those verifying the scaling $\eta \sim R^{2}Re^{1/2}$ have a capture mode of type `direct interception', as qualitatively described by definition. The fact that these particles are intercepted within the boundary layer led us to think of direct interception as a consequence of the particle diameter being smaller compared to the boundary layer. How small should it be? The answer we provide in this paper is 70\% of the boundary layer thickness. We recall that this value was obtained assuming that the boundary layer thickness equals exactly the inverse of the square root of the Reynolds number $\delta/D = 1/Re^{1/2}$. Formally, nonetheless, we should consider a numerator $f(\theta)$ varying with the angular position $\delta/D = f(\theta)/Re^{1/2}$ because the thickness is not constant along the cylinder's arc of circle. According to the numerical simulations of Haugen and Kragset,\cite{haugen_particle_2010} the function $f(\theta)$ does not vary much around the cylinder: it is equal to 1.7 at $\theta = 10\degree$, drops down to 0.45 at $\theta = 55\degree$, then climbs up to 0.8 at $\theta = 90\degree$. Thus, it is fair to take unity as a representative value. To avoid any confusion, we stress that the quantitative definition of direct interception we propose should read verbatim $\deep/\delta = RRe^{1/2} < 0.7$.

This formulation also shows that the validity range of the scaling $\eta \sim R^{2}Re^{1/2}$ depends on the Reynolds number and diameter ratio altogether through the product $RRe^{1/2}$, and not only on the Reynolds number as described in Weber and Paddock \cite{weber_interceptional_1983} and Espinosa-Gayosso \textit{et al.}\cite{espinosa-gayosso_particle_2013}. It makes sense why they got distinct Reynolds ranges: the validity range was $1 \le Re \le 200$ for the former, and $200 \le Re < 1000$ for the latter.
Haugen and Kragset \cite{haugen_particle_2010} were more careful by stating a validity range in terms of the Stokes number which regroups $R$ and $Re$, i.e. $Stk \sim R^{2}Re$. As such, they indirectly agree with our idea about comparing the sizes of the particle and boundary layer given that $Stk \sim \left(RRe^{1/2}\right)^{2} \sim (\deep/\delta)^{2}$. Having said that, we believe that reasoning in terms of the ratio $\deep/\delta$ is more insightful since it is a geometrical comparison between sizes, whereas the Stokes number, by definition, is a comparison between timescales.

As for the model of particle advection, we included the added mass force to the particle momentum balance, which was ignored in the work of Haugen and Kragset.\cite{haugen_particle_2010} We considered an always constant mass coefficient (equal to 1/2 for spherical particles). This assumption becomes invalid near the cylinder, though, because particles would experience a repulsive force as they approach the cylinder. This issue can be rectified by entering an expression of the added mass coefficient varying with the distance to the cylinder wall,\cite{brennen_review_1982} or model the repulsive force, as did Béguin \textit{et al.},\cite{beguin_void_2016} then add it in the force balance. Moreover, since particles are constantly swimming or undergoing Brownian motion,\cite{guasto_fluid_2012} a realistic effect could be depicted by including, for instance, a random displacement proportional to the diffusion coefficient while integrating the momentum equation.

Also we should mention that we simulated two-dimensional DNS even for Reynolds numbers $Re \ge~200$, where vortex shedding becomes three-dimensional. Espinosa-Gayosso \textit{et al.}\cite{espinosa-gayosso_particle_2013} emphasised that running two-dimensional simulations in this regime returns erroneous results, and three-dimensional DNS is the right choice.
It is believable that three-dimensional DNS could lead to some circles ($\bigcirc$) in the region $Re \ge 200$ in Fig.~\ref{fig:phase_diagram} to be converted into crosses ($\times$) and vice versa. We recognise that the separatrix $R^{*} = 0.7/\sqrt{Re}$ between direct interception and inertial impaction might be altered for capture events in flows at $Re \ge 200$.
Notwithstanding, there is room for this separatrix to hold even in three-dimensional flows, because the streamlines upstream from the cylinder, which determine the scaling as we have seen in the theoretical derivation, are left unaffected by three-dimensional artifacts. Three-dimensional variability in the flow starts slightly at the rear of the cylinder and becomes important only downstream from it. We leave the door open for future works to inspect the robustness of the criterion $R^{*} = 0.7/\sqrt{Re}$ in three-dimensional regime.

Finally, when the particle diameter is larger than 70\% of the boundary layer thickness, the dominant capture mode is the inertial impaction, and the scaling of the capture efficiency has an exponent in the Reynolds number greater than 1/2 (more precisely linear in the Reynolds number $\eta \sim Re$). From this perspective, we can explain the scaling $\eta \sim Re^{0.7}$ in the experiments of Palmer \textit{et al.} \cite{palmer_observations_2004} and simulations of Haugen and Kragset \cite{haugen_particle_2010} by the fact that the capture mode was transitioning to an inertial impaction. In this way, our definition wipes out the apparent contradiction between results in prior studies.

Despite the limitations mentioned above, the classification of capture modes by comparison of the particle diameter with the boundary layer thickness is general and practical in research pertaining to particle capture. It provides a clear cut and quantitative limit to distinguish between direct interception and inertial impaction, which is essential to reconcile the pretended discrepancies of results regarding capture efficiency in the literature. We hope this classification will be useful and serve a wide range of studies in the field of particle filtering.

\begin{acknowledgments}
We acknowledge funding from the Simulation-Based Engineering Science (SBES) program of the National Science and Engineering Research Council of Canada (NSERC), as well as from Discovery Grants Nos. RGPIN-2019-07072 and RGPIN-2019-05335.
\end{acknowledgments}

\appendix

\section{Mathematical method to determine the separatrix}
\label{app:mathematical}

In section \ref{sec:modes} we defined direct interception as the capture mode of any particle of diameter ratio $R$ in a flow of Reynolds number $Re$ verifying the following inequality
\begin{equation}
\frac{\deep}{\delta} = R Re^{1/2} < M.
\end{equation}
Graphically, we found that the constant $M = 0.7$ yields the right separatrix $R^{*} = 0.7/\sqrt{Re}$ that splits well the numerical data in the phase diagram in Fig.~\ref{fig:phase_diagram}. In this appendix we propose a more rigorous way to find the value $M = 0.7$.

Let $\scrD ir$ be the set of particle advection cases where the capture mode is a direct interception
\begin{equation}
\scrD ir = \left\{ (R, Re) \in [0.008, 0.1] \times [3, 300],\ 
\frac{\deep}{\delta} = R Re^{1/2} < M
\right\},
\end{equation}
which can also be rewritten as
\begin{equation}
\scrD ir = \left\{ R \in [0.008, 0.1], \
\sqrt{Re} < \min \left( \sqrt{300}, \frac{M}{R} \right)
\right\},
\end{equation}

Our analysis of the trajectories in Fig.~\ref{fig:boundary_layer} revealed that captured particles following the scaling $\eta \sim R^{2}Re^{1/2}$ experience a direct interception. Yet, we have seen from the curve fitting in Fig.~\ref{fig:eta_vs_R2sqrtRe} that a subset of particles verifying this scaling is
\begin{align}
\scrS ub &= \left\{ (R, Re) \in [0.008, 0.1] \times [3, 300],\ 
R^{2}\sqrt{Re} < 0.028
\right\} \nonumber \\
&= \left\{ R \in [0.008, 0.1],\ 
\sqrt{Re} < \min \left( \sqrt{300}, \frac{0.028}{R^{2}} \right)
\right\}.
\end{align}

Thus, we should have
\begin{equation}
\scrS ub \subset \scrD ir.
\end{equation}

It implies that the upper bound of the first set is necessarily smaller than the upper bound of the second set
\begin{equation}
\forall R \in [0.008, 0.1],\ 
\min \left( \sqrt{300}, \frac{0.028}{R^{2}} \right)
< \min \left( \sqrt{300}, \frac{M}{R} \right),
\end{equation} 
or equivalently, by multiplying by $R$ in both sides,
\begin{equation}
\forall R \in [0.008, 0.1],\ 
\min \left( \sqrt{300}R, \frac{0.028}{R} \right)
< \min \left( \sqrt{300}R, M \right).
\label{eq:inequality_two_functions}
\end{equation} 

The left-hand side function in \eqref{eq:inequality_two_functions} has a maximum at $R_{0}^{*} = \sqrt{0.028/\sqrt{300}} \approx 0.04$. Then $M$ is taken as the smallest constant verifying the inequality \eqref{eq:inequality_two_functions} for all particles considered, which is
\begin{equation}
M = \sqrt{300} R_{0}^{*} \approx 0.70.
\end{equation}

\section*{Data availability}
The data that support the findings of this study are available from the corresponding author upon reasonable request.

\bibliography{msc_biblio}

\end{document}